# An Electrochemical Appraisal of Coffee Qualities

Robin E. Bumbaugh[a,†], Doran L. Pennington[a,†], Lena C. Wehn[a], Elias J. Rheingold[a], Joshua R. Williams[b], Benjamín J. Alemán[c], Christopher H. Hendon[a]*

[a] Department of Chemistry and Biochemistry, University of Oregon, Eugene, OR, 97403, USA *email: chendon@uoregon.edu

[b] Department of Chemistry, Drexel University, Philadelphia, PA, 19104, USA

[c] Department of Physics, University of Oregon, Eugene, OR, 97403, USA

[†] These authors contributed equally.

## Abstract

Despite coffee's popularity, there are no quantitative methods to measure a chemical property of a black coffee drink *in situ* and relate it to a flavor experience. Here we show that cyclic voltammetry can be used in coffee without any additional sample preparation to directly measure the strength of a coffee beverage and, separately, how dark the coffee has been roasted; these two properties are implicated in the sensory profile of the beverage. We show that the current passed for the cathodic features that precede hydrogen evolution are linearly related to beverage strength. The same features are suppressed with subsequent cycling, and we show that the magnitude of suppression is directly related to roast color, which dictates the ensemble chemical composition and thus flavor of the beverage. Together, this voltametric analysis decouples beverage strength from roast color and offers a new strategy for rapidly assessing chemical properties of coffee that correlate to flavor.

**Introduction**

Since the 1950s, the coffee industry has sought quantitative methods to assess beverage qualities beyond those informed by sensory panels. In the meantime, a litany of research on the topic has revealed that beverage concentration[1], [2] and roasted bean color[3], [4] are the two primary and independent factors that dictate the sensory perception of coffee. Bean color is readily determined by spectrophotometry[5], [6], [7], while the most widely used technique to measure concentration of solvated coffee relates the refractive index of the beverage[8], [9] to an effective concentration through an empirically derived polynomial. The refractive index method reports an approximated total dissolved solids (TDS) ratio of the mass of dissolved coffee to the mass of the beverage and informs beverage strength[10]. From %TDS, the efficiency of the extraction process can then be calculated by taking the mass ratio of the dissolved solids to the dose of ground coffee (*i.e.*, extraction yield[11]). The industry has coalesced around the phenomenological observation that dissolution of ~20% of the dry mass, yielding a beverage ~1.35 %TDS, generally produces an enjoyable cup of filter coffee[12], [13]. To date, measurements based on refractive index remain the industry standard. However, the refractive indices of solutions are dependent on the identity of the solute. Glucose, for example, will have the same refractive index at 2% w/w as a solution of ethanol that is twice as concentrated, making refractive index measurements unable to identify chemical differences in coffee that give rise to dissimilar sensory outcomes[14], [15], [16], [17], [18], [19]. Given the composition of roasted coffee primarily depends on roast color, the existing approach quantitative approach cannot discern differences between light-roasted and dark-roasted coffees that achieve the same refractive index, let alone higher-fidelity chemical differences achieved in the same coffee using modified brewing parameters or roast profiles. A rapid quantitative method that is sensitive to compositional information beyond %TDS remains a major target for the industry.

There exist numerous analytical techniques that provide both qualitative and quantitative information about chemical composition, with the gold standard being liquid or gas chromatography coupled with mass spectrometry for soluble and volatile compounds, respectively. Besides the obvious challenge of identifying specific compounds among the thousands observed in coffee that give rise to a measurable sensory experience, these techniques also suffer from slow run times, laborious sample preparation, and high associated costs while yielding limited predictive insights. Instead, some research groups have used electrochemical techniques to measure the concentrations of common molecules in solution[20], [21]. Electroanalytical techniques measure the amount of current passed between electrodes immersed in the solution at voltages where solvated molecules undergo redox reactions; the measured current is directly proportional to the local concentration of the molecule. While this approach is generally sensitive enough to accurately measure the concentration of caffeine[22], various chlorogenic acids (CGAs)[23], and other molecules[24], previous reports have omitted beverage strength in their analyses, precluding the development of a technique which relates preparation variables (*e.g.* mass of coffee, mass of water, grind setting, water temperature and pressure, contact time, roast color etc.) to the resultant chemical composition of the liquid.

Herein, we report an advance in coffee quality analysis that harnesses *in-situ* changes in the protonic electrochemical response (*e.g.*, hydrogen underpotential deposition, $H_{UPD}$) in cyclic voltammetry (CV) measurements of liquid coffee. By sampling features in the electrochemical response that are affected by the ensemble chemical composition of the coffee rather than measuring the concentrations of individual molecules, this approach remarkably captures quantitative information about both roast color and beverage strength. These two properties drive the sensory profile of the beverage, thereby allowing this analytical technique to vastly exceed the insights provided by refractive index measurements, and provide a highly novel quality control technique.

**Results and Discussion**

Initially, we sought to directly measure the concentration of caffeine ($E_{ox}$ = ~1.4 V vs. Ag/AgCl[25]), CGAs ($E_{ox}$ = ~0.2-0.5 V vs Ag/AgCl, depending on the isomer[26]), and other redox-active species in undiluted coffee samples to study their dependence on conventional brew parameters. However, caffeine and CGAs form an aggregate at concentrations typical of brewed filter coffee[27], terminally impacting their redox activity — they only become electrochemically well-resolved in acidified dilute solutions with added electrolyte. Literature examples dilute to below 0.02 %TDS[28], nearly two orders of magnitude weaker than filter coffee. Other groups have shown that by adulterating coffee with either CGA or caffeine, both boron-doped diamond (BDD) and glassy carbon (GC) electrodes can provide molecule-specific information[29], [30], [31], [32], [33], [34], [35], [36], [37]. Yet, these papers required laborious sample preparations, in opposition of both our and the industry goal of measuring features of as-consumed coffee. Thus, we first ascertained the redox landscape in filter-strength coffee by taking CV measurements using BDD, GC, and Pt working electrodes across the electrochemical window of the beverage, **Figure 1a**.

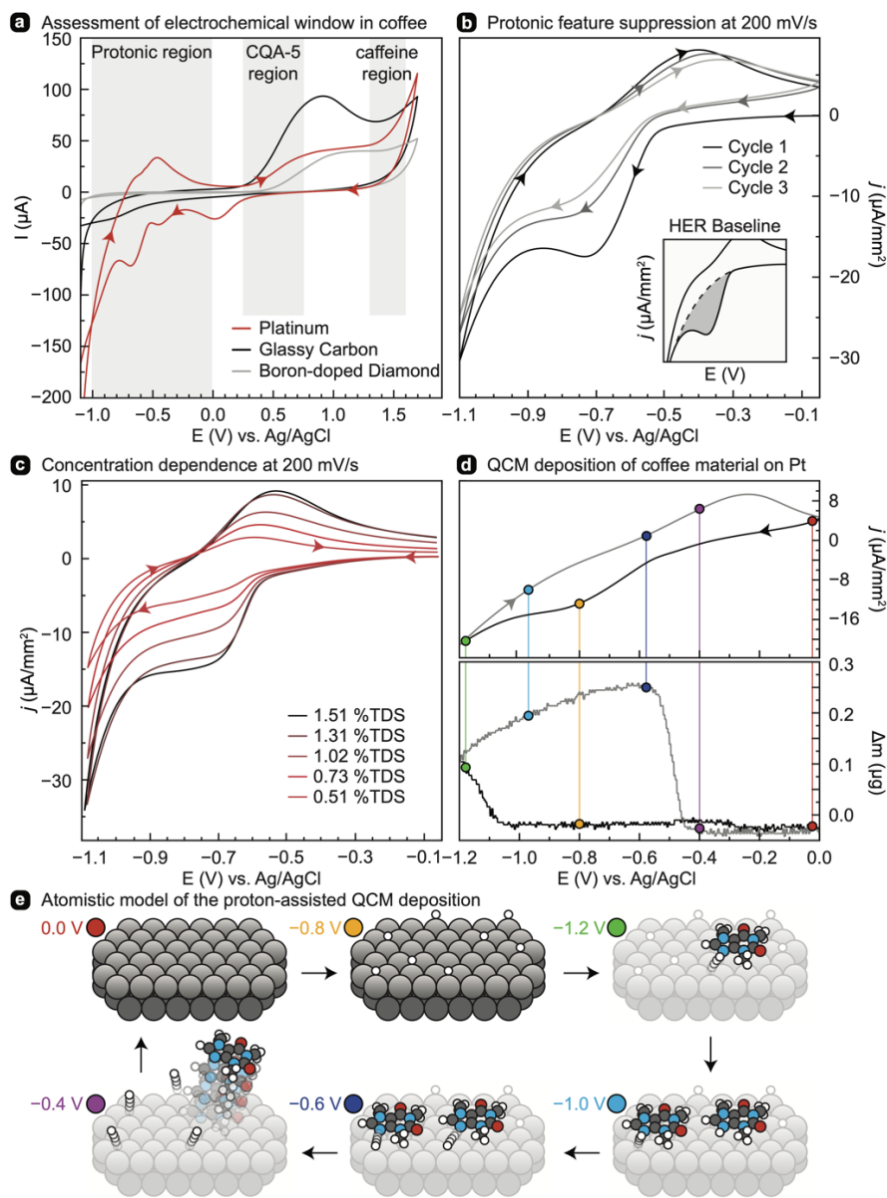

**Figure 1. Assessing the voltametric features present in coffee extracts. a)** Cyclic voltammetry performed at 200 mV/s using boron-doped diamond, glassy carbon, and platinum working electrodes. Caffeine and chlorogenic acid potential ranges studied by other groups are highlighted. Here, we focus on the protonic features ($H_{UPD}$ and acid redox) region. **b)** These features are suppressed with subsequent CV cycling due to deposition of coffee material on the surface of the electrode, and the total charge can be extracted by subtracting the hydrogen evolution reaction (HER) background. **c)** The current depends on %TDS of the brewed coffee because the protonic and organic concentration scales with %TDS. **d)** Scanning anodically results in mass accumulation on the working electrode, leading to electrode fouling. **e)** However, the mechanism of mass accumulation is likely proton-assisted, given that appreciable mass does not deposit until the surface has accumulated a critical H-atom concentration.

As-consumed filter coffee extracts are sufficiently conductive (bulk conductivities are typically 3.0 ± 0.2 mS/cm) for direct electrochemical analysis without the addition of a supporting electrolyte and are self-buffered to pH of ~4.8 – 5.9 depending on the distribution of compounds in the coffee and the water composition[38], [39], [40]. Even after performing bulk electrolysis at oxygen evolution potentials for two minutes, the pH of the solution remains numerically identical, reinforcing the significant buffering capacity of brewed coffee. Yet, despite the plethora of molecules in coffee extracts, the CV response of a Pt working electrode in 1.56 %TDS coffee is consistent with that of dirty acidic water, **Figure 1a**[41], [42], [43], [44], [45], [46].

The cathodic Faradaic features map to the response expected for protonic reactions with the Pt surface (e.g., $H_{UPD}$), followed by $H_2$ evolution at more negative potentials. At positive potentials, OH adsorption and eventual $O_2$ evolution are also evident. To ascertain whether the anodic feature at –0.6 V is linked with the cathodic features in a reversible redox couple, we probed the scan rate dependence, **Figure S1**. The linear dependence of peak current on the square root of the scan rate for both features demonstrates the diffusion-controlled nature of the redox events, **Figure S2**, and the lack of an increase in the peak potential separation with increasing scan rate indicates Nernstian behavior, **Figure S3**. However, the large peak separation of ~200 mV at all scan rates suggests that the reversibility of the redox couple is obfuscated by sluggish kinetics.

The same Pt surface sites that adsorb $H^+$ and $OH^-$ are also able to adsorb other molecules in solution. In the case of oxidative cycling, some impurities in water compete for the Pt surface, resulting in reduced current with subsequent cycling due to a decrease in the accessible surface area. Given that Pt is known to interact with caffeine and other molecules in coffee[47], [48], we expected to see a decrease in exchange current density with sequential cycling. When scanning from 0 to –1.0 V, the $H_{UPD}$ and protonic features ($E_{pc}$ = –0.4 and –0.7 V, respectively) smear together and current decreases by ~34% from CV scan 1 to 2 and ~18% from scan 2 to 3, **Figure 1b**.[49] In pH 7 water purified by reverse osmosis the same features are not observed, **Figure S4**, suggesting that the response is due to protonic chemical steps associated with the coffee and not the water.

Further experiments were run to ensure that these features mapped to $H_{UPD}$/weak acid reduction and its suppression by coffee molecules, rather than fluctuations in dissolved $O_2$ and other spurious effects, **Figure S5**. Since the integral of the current density depends on the activity of $H^+$, the $H_{UPD}$ and acid features indirectly provide ensemble insights into the families of molecules in coffee that function as $H^+$ donors and acceptors, the concentrations of which should depend on roast color, brewing parameters, brew water composition, and so forth. Some data in support of this hypothesis is that $H_{UPD}$/acid reduction current density decreases with decreasing coffee concentration, **Figure 1c**, due to the diminished concentration of available $H^+$. Because the feature is concentration dependent, there are also fewer organic molecules competing to bind to the surface of the electrode. As we will show later in **Figure 3**, the integral of the charge current density of this feature linearly maps to %TDS. Perhaps this is a surprising result, given a single acidic feature should not necessarily depend on ensemble concentration.

To further support our proposed mechanism that protonic chemistry is being suppressed by adsorbed coffee material, we performed CV measurements using an electrochemical quartz crystal microbalance (QCM) with a Pt working electrode scanning at 50 mV/s, **Figure 1d**. Scanning cathodically, appreciable mass begins to accumulate on the electrode at potentials more negative than $H_{UPD}$ — the balance is insensitive to surface H-adsorption but can detect larger molecule accumulation. There is a delayed onset, which attribute to a proton-assisted adsorption of Brønsted-basic species like caffeine, following a general mechanism presented in **Figure 1e**. Upon scanning anodically from –1.2 to 0.0 V, mass continues to accumulate until the potential exceeds –0.5 V, when the electrode liberates most of the adsorbed organic material and protons back into solution. The response is reminiscent of a kinetic trap, where the surface assembly forms in kinetically-favored conditions[50]. To ensure that the process was proton assisted, we also scanned from –1.2 to 0.0 V without first scanning cathodically and detected no mass accumulation, indicating that there must be an appreciable concentration of protons on the surface before larger organic species begin to accumulate, **Figure S6**.

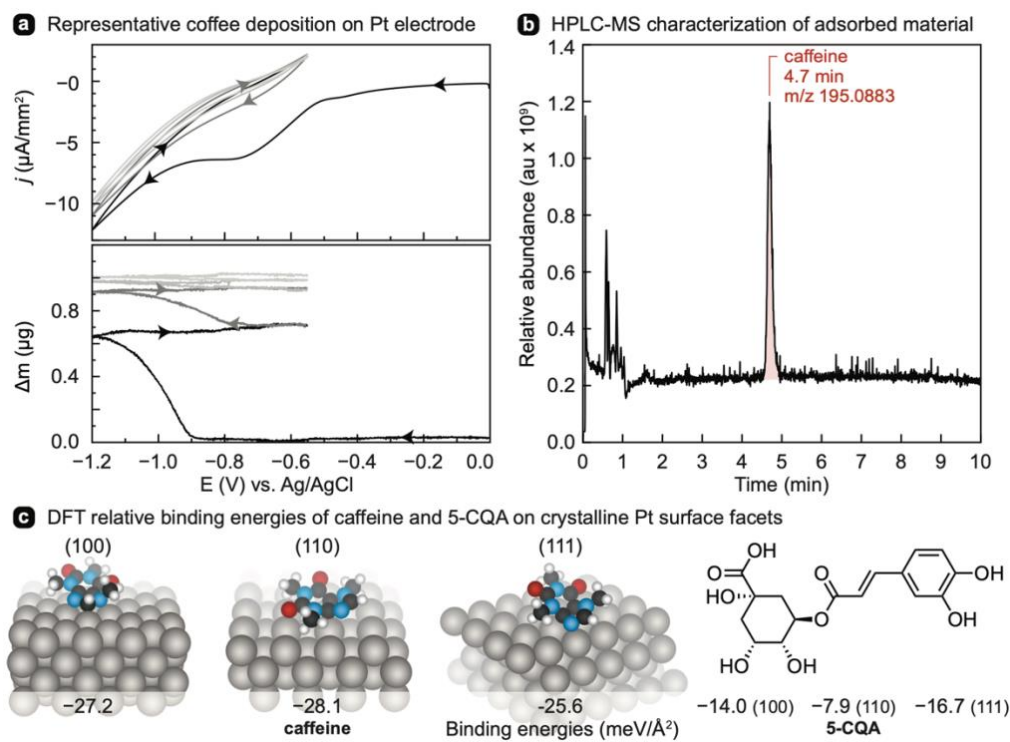

**Figure 2. Identification of surface adsorbates at negative applied potentials. a)** Cyclic voltammetry performed at 200 mV/s sampling potentials more negative than –0.5 V, to ensure our accumulated mass is maximized rather than liberated back into solution, per **Figure 1d**. The first four cycles of the second run are presented. **b)** Combining the extracts from the surface of the Pt electrode sampled 100 times in four separate runs, we can detect the presence of caffeine and quantify it using the calibration curve presented in **Figure S7**. **c)** The adsorption of caffeine and 5-caffeoylquinic acid are favored on all clear crystal facets of Pt according to density functional theory simulations, suggesting that a collection of organic molecules responsive to roast color and brewing parameters adsorb to the electrode surface.

While the total charge passed maps linearly to concentration for any particular coffee, coffees from different origins, processed in dissimilar ways, and roasted to different colors may show major differences in the emergence and suppression of the reductive features. Before we can probe coffee-related variables, we must first ensure that the suppression of the convoluted redox feature at –0.55 V depends on molecules likely found in all coffees. To determine the molecular identities of the adsorbed compounds, we developed the CV cycle shown in **Figure 2a** to maximize mass accumulation on the Pt working electrode for further characterization. We were able to obtain sufficient coffee material on a Pt-mesh electrode surface by cycling 100 times from –0.55 to –1.2 V, **Figure 2a,** followed by solvating the adsorbed material in a sonicated 4 mL bath of 80/20 water/acetonitrile (v/v), and repeating four times. The adsorbates could then be separated and characterized using high performance liquid chromatography coupled with high-resolution mass spectrometry (HPLC-MS), **Figure 2b**. We found that caffeine had adsorbed in quantifiable concentration. Our combined samples yielded 7.8 ± 0.1 mg/kg caffeine, suggesting that at least one component of the accumulated mass that causes the current to decay over successive CV cycles originates from a molecule common to all coffees, and that our 4 mL solution contained approximately 300 µg of caffeine, or ~0.4% of the total caffeine in an average 180 mL cup of filter coffee[51], [52], [53] (see **Figure S7,S8**). That is, each 100-cycle CV presented in **Figure 2a** scavenged approximately 0.1% of the available caffeine in the cup, as well as other molecules.

Given that a similar electrochemical approach has been used to quantify caffeine content in highly dilute coffee samples through oxidation [30], and that bulk Pt also is known to adsorb organic material [41], we were somewhat unsurprised to see caffeine in the chromatogram. However, there are of course some coffees that have been decaffeinated, prompting us to resolve whether this mass deposition can be attributed to additional adsorbates beyond caffeine. Because other adsorbates were not in sufficient concentration to quantify with HPLC-MS, we instead turned to density functional theory (DFT) paired with molecular dynamics simulations (MD) to model the Pt surface adsorption energies for caffeine as well as 5-caffeoylquinic acid (5-CQA), an abundant CGA isomer found in all coffees with a concentration that depends strongly on roast parameters [51], [54], [55], **Figure 2c**. The inclusion of caffeine in the DFT study serves as a control to validate the model, and the inclusion of 3 low-index Pt surfaces, (100), (110), and (111), examines the possibility of preferential binding to certain facets of the polycrystalline Pt electrode used for the electrochemical measurements. No solvent molecules were included in the model for computational efficiency, but similar approaches for Pt surface adsorption studies find excellent agreement to experiment.[76]

For each combination of molecule and surface, we obtained several local-minimum configurations from the MD trajectories and then ran DFT structural optimizations to obtain equilibrium energies; the difference in energy between the molecule-surface complex and the isolated molecule and surface defines the adsorption energy, with a more negative value representing stronger binding. The adsorption of caffeine to three facets of Pt revealed each facet stably binds caffeine as expected, with a slight preference for the (110) surface. Our calculations reveal that 5-CQA also forms stable complexes with the three Pt surfaces, albeit with a weaker interaction than caffeine, and prefers the (111) surface. Interestingly, 5-CQA binds least strongly to the (110) surface, indicating that unique crystal facets of the Pt electrode may offer some degree of selectivity for adsorbed species, which in turn may prove useful in follow-up work that harnesses the kinetics of adsorption to parse flavors. Together, the DFT models instruct that the suppression of the $H_{UPD}$ and protonic features likely capture the ensemble of various organic adsorbates binding to the electrode surface, providing information about the beverage concentration, as well as roast color, since roast will dictate the amount of 5-CQA (and caffeine [56]) in the cup.

Since the Faradaic protonic features depend on coffee concentration, we sought to elucidate how roast color (and subsequent chemical composition) impacts the redox response. To do so, we roasted a representative specialty coffee sourced from Colombia, following the profiles shown in **Figure 3a**. The "light roast" profile was then systematically extended by 30 seconds and in final temperature by 1.11 °C (2 °F), to achieve four progressively darker coffees. A final sample was prepared by extending the fourth roast by a further 60 seconds and 2.22 °C (4 °F) to ensure that the roast colors covered a broad and reasonable range of Agtron values, see Methods. The roast profiles yielded coffees ranging from 75.8 (light) to 55.7 (dark). Each roasted sample was rested for 7 days to allow for $CO_2$ off-gassing[57] and then brewed using the Specialty Coffee Association cupping protocol.[58]

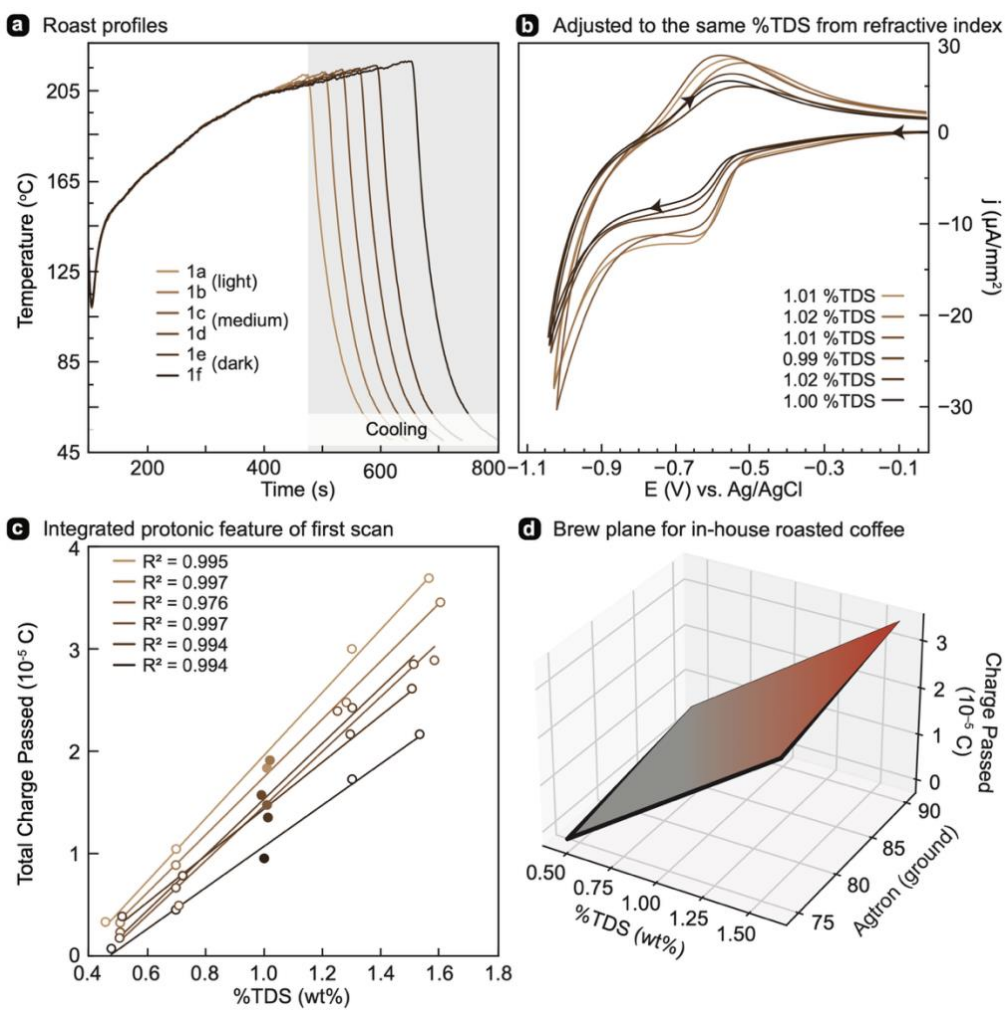

**Figure 3. Relating roast profile and beverage strength to the cathodic feature at –0.6 V. a)** The measured temperature profiles used to generate six systematically darker roasts. **b)** The first cycle CV response of these coffees brewed and diluted to 1.00 ± 0.02 %TDS. **c)** Subtracting the background contribution due to the onset of hydrogen evolution (**Figure S9**), the total charge passed is linear with %TDS with a slope dependent on roast color. Darker roasts more effectively suppress the cathodic features, likely because they contain more solvated oxidized molecules formed during roast. These compounds evidently tend to bind more strongly to Pt. **d)** Plotting the total charge and %TDS against the ground coffee Agtron value (roast color), a 3D planar

relationship is recovered. Error bars (±0.01 %TDS, ±0.1×10$^{-5}$ C) are omitted for clarity but can be found in **Figure S10**.

To isolate roast dependence, some sample preparation is required. By diluting the extracts to 1.00 ± 0.02 %TDS, electrochemical differences that depend on roast color were noted in the voltammograms, **Figure 3b**. The lightest-roast coffee passes ~50% more charge than the darker analogues at the same refractive index strength. While %TDS remains fixed, the electrochemical assessment of the cathodic feature depends on roast, a critical parameter that dictates the sensory experience (*i.e.*, dark-roasted coffees taste "dark"). By integrating the exchange current density and adjusting for scan rate, we can extract the total charge passed for each coffee at any arbitrary concentration, **Figure 3c**. Here, each roast shows a strong linear fit between %TDS and electrochemical charge passed. Increasingly darker roasts yield both a shallower gradient and less total charge passed than the lighter-roast counterparts. The diminished current may be expected from the fact that darker coffees tend to have mildly elevated pH[59], [60], and that dark roasts have less water-soluble material contained within, thereby preferentially depositing organic material onto the electrode. The rate of suppression of these redox waves depends on composition (see **Figure S11-S16**). Together, we can conclude that the while the feature is indirectly related to coffee composition, it is sensitive to differences in roast-derived species.

The %TDS can then be plotted along with the coffee Agtron color (both whole bean and ground show the same trend, see **Figure S22**) and total charge passed in the $H_{UPD}$ region to yield a plane with unique values of integrated charge for all combinations of %TDS and Agtron color, **Figure 3d**. While the shape of the plane may be coffee-specific, the fact that roast color maps linearly to charge allows for a wealth of applications. For example, a series of simple CV measurements on progressively more dilute coffee will allow a roaster to rapidly construct a quality control calibration curve, enabling quantitative comparisons of separate batches of the same coffee roasted to the same color. In a roastery quality control setting, it is possible that batches of roasted coffee could have similar whole bean colors but different flavor profiles, which may be detectable electrochemically as well as on the cupping table.

As a demonstration of the analytical capabilities of this approach, we sourced four batches of the same coffee roasted to a target same whole-bean Agtron color (93.0 ± 1.0) from a specialty roaster (Colonna, Bath UK). Among those four batches, one was rejected by the roaster because it did not meet the color tolerance (the coffee was too light, 98.9), and it consequently exhibited undesirable flavors during their sensory quality control. All coffee roasters have financial motivations to accept their roast batches: consumers can be quite confident that the coffee they drink is not a rejected sample. We wondered whether we could discern an electrochemical difference between the accepted and rejected samples, in essence performing at least as well as the human tongue.

The roaster provided these samples to us labeled 1, 2, 3, and 4, and did not specify which was rejected. Single-blind, we prepared each sample in triplicate in a randomized order and brewed them according to the Specialty Coffee Association cupping technique. The samples were undiluted to allow %TDS to vary thereby introducing realistic error to our samples. After performing CV measurements on the samples in yet another randomized order, statistical analysis was performed using Tukey's honestly significant difference analysis of variance. The analysis revealed that the measured %TDS of each coffee was statistically identical (**Figure 4**);

neither color analysis nor refractive index could tell these coffees apart. However, examination of the current passed in scan one revealed that sample 1 was statistically dissimilar to the other coffees ($p < 0.0039$). The same was found for scans two and three. However, examining the difference in current passed between scan one and two, normalized by the initial current in scan one (*i.e.*, $(A_1–A_2)/A_1$) revealed that the rate of fouling of the electrodes were the same for all coffees ($p > 0.832$). That is, the rate of electrode fouling depends on concentration, but the charge depends on composition. By consulting with the roaster, sample 1 was revealed to be the rejected batch. Furthermore, the acceptable batches all lied within the same statistical class, demonstrating alignment between this electrochemical analysis and sensory profiling.

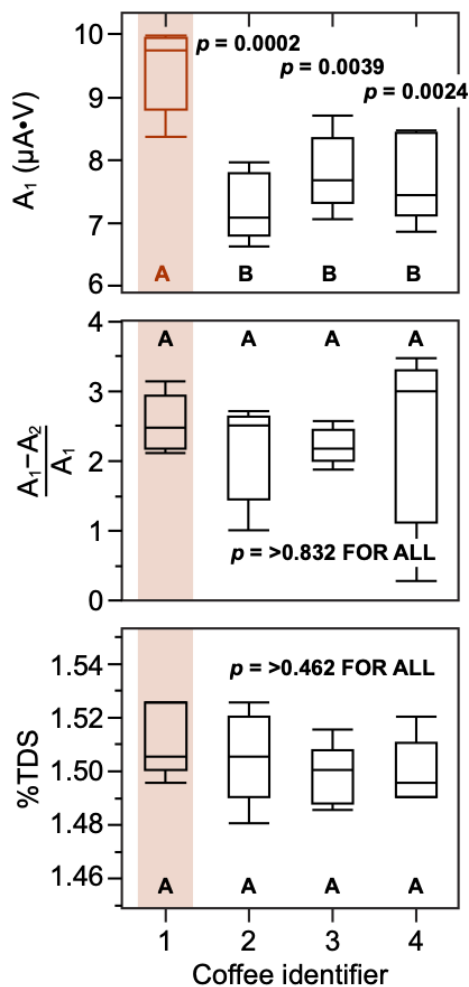

**Figure 4. Electrochemical quality control at the roastery.** A single-blind assessment of four coffees roasted to the same specification, but one (sample 1) exhibits negative flavors due to minor variation in roast parameters. Tukey's honestly significant difference ANOVA results reveal the measured %TDS was statistically identical across all four samples (connecting letters A class); TDS differences (insignificant) ranged from 0.002 ± 0.02 to 0.03 ± 0.02, with $p = 0.999$ to $p = 0.462$, respectively. The background-subtracted integrated cathodic area at –0.6 V for the first cycle, $A_1$, was split into two distinct classes, A and B, with sample 1 in class A and samples 2, 3,

and 4 in class B. The differences between classes A and B were 2.19 ± 0.40 µA·V (1-2; $p$ = 0.0002), 1.73 ± 0.40 µA·V (1-4; $p$ = 0.0024), and 1.64 ± 0.40 µA·V (1-3; $p$ = 0.0039). Differences between batches in the B class (insignificant) ranged from 0.09 to 0.55 µA·V ($p$ = 0.996 to $p$ = 0.529, respectively). The fouling was captured by normalizing the differences in charge passed between scans one and two, where all four samples were found to exhibit indistinguishable decreases in current with subsequent scans. A comprehensive comparison is presented in **Figure S20** and **S21**.

Considering the findings in this paper, there is obvious upside to electrochemical assays of coffee extracts. Coffees can now be quality controlled and distinguished by their charge response — which is linked to the collection of molecules competing for the Pt surface — rather than the ensemble effect on refractive index. Given we know that the response depends on the concentration of $H^+$ (correlated with perceived acidity)[61] but is suppressed by roast-dependent molecules in coffee like caffeine (bitter)[62] and chlorogenic acid (astringent[63], sour[64]), the electrochemical technique implicitly provides insights about flavor. Finally, this approach is extremely sensitive to composition and hence can be used as a highly effective sensor which may find use in achieving desirable blends, detecting differences in seasonal crops, and resolving other high fidelity coffee variables.

**Conclusion**

We have demonstrated that CV measurements performed in the cathodic region of the water window of brewed coffee produce a response aligned with the deposition of $H^+$ on the Pt surface before forming $H_2$. Since the charge passed during $H_{UPD}$ depends on the number of available surface sites on the Pt electrode, which is a function of the ensemble organic composition of the solution as well as its pH, these electrochemical features serve as a highly useful and sensitive measure of ensemble chemical composition in coffee samples. By performing a series of careful experiments, we show that the Pt surface also accumulates coffee material, leading to a reduction in charge passed with subsequent cycling, and identify that at least some fraction of this mass is associated with caffeine and other molecules in coffee, but not the brewing water itself. DFT simulations further instruct that the adsorption should generally have some dependence on molecular composition, and hence different coffees should suppress the cathodic feature to different extents. The most straightforward method to assess this was to simply prepare different roast levels of the same coffee. We demonstrated that in that series, coffees exhibit a linear relationship between beverage strength and total charge passed in the $H_{UPD}$ region, and that the rate of charge suppression with change in %TDS depends on roast color. However, the technique is even more sensitive, as validated by matching human sensory discrimination (performed at a roastery during their quality control assessment). Our measurements successfully assign statistical classifiers to acceptable and unacceptable batches of roasted coffee with the same physical attributes. This suggests that the electrochemical method is sensitive to composition and together offers a new and reliable method to measure a critical aspect of coffee composition, as well as beverage strength, simultaneously.

**Methods**

*Sample Preparation*

Green coffee was roasted in-house as described in the *Roasting* section. These roasts canvassed industrially relevant light, medium and dark roasts (spanning whole bean Agtron values of 75.8 (lightest) to 55.7 (darkest), **Table S1**). The coffee was allowed to rest for 7 days before being ground with a Mahlkönig EK-43 grinder (particle size distributions are presented in **Figure S17**) and brewed via cupping method with a brew ratio of 1:13.5 using 93°C water (filtered Eugene tap water using a Pentair Everpure Conserv 75E Reverse Osmosis system with remineralization to 30 mg/L $CaCO_3$). The coffee was allowed to contact the water for 4 minutes, without agitation or stirring. The samples were then filtered through a V60 filter paper to arrest brewing and allowed to come to room temperature (21.6 °C). The samples were stirred before each aliquot was taken to ensure homogeneity. The cumulative concentration of solvated coffee material (total dissolved solids as a mass percentage) was calculated from the measured refractive index following a literature procedure[65].

*Electrochemical Measurements*

A three-electrode electrochemical setup was employed for all voltammetric measurements. Cyclic voltammetry (CV) was performed using a Gamry 1010B potentiostat controlling Pt disk working (surface area, 0.0314 $cm^2$) and Pt wire counter electrodes, and an Ag/AgCl (sat. KCl) reference electrode, purchased from CH Instruments. Prior to data collection, the working and counter electrodes were polished using a modified literature approach[66] (10 scans at 300 mV/s from −0.3 to 1.0 V in 50 mM $H_2SO_4$, until key $H_2SO_4$ redox features overlaid, **Figure S18**). Signal-to-noise was then assessed through comparison between the brew water, the cleaning solution, and the coffee samples, **Figure S19**. Measurements were made directly on as-brewed coffee, with no supporting electrolyte or buffer. The solution specific conductance of the coffee was measured using a Fisher Scientific Traceable conductivity probe calibrated with standard saturated solutions of KCl. The pH of each solution was noted in **Table S1**. Since most coffee is a solution of weak acids, the solutions are self-buffering to a pH of around ~5.0, aligned with previous literature reports[39].

*Particle Size Distribution*

Particles created by the EK-43 were measured using a Malvern Mastersizer 2000 laser diffraction system. The solid samples were dispersed into an airstream (2.0 bar) of compressed breathing-quality air and passed through a He-Ne laser at 633 nm and solid-state blue light at 466 nm. Ground coffee samples were run in triplicate and the average particle size distributions are plotted in **Figure S17**.

*Quartz Crystal Microbalance*

A QCM200 Quartz Crystal Microbalance equipped with a crystal holder was used to perform all quartz crystal microbalance (QCM) experiments. Pt-coated liquid plating 5 MHz quartz crystal electrodes with a Ti adhesion layer from Fil-Tech were used. The crystal holder ensures that only one side of the Pt coated electrode is accessible to the solution and the controller reads the associated frequency. A Biologic SP-300 potentiostat with a Pt wire counter electrode and Ag/AgCl (sat. KCl) reference electrode was used for collecting CV measurements corresponding with QCM data. The QCM frequency data was converted to mass values via the Sauerbrey equation as follows:

$$\Delta f = \frac{-C_f \Delta m}{A}$$

Where $f$ is the frequency as read from the QCM controller, $C_f$ is the calibration constant which is given to be 56.6 Hz·cm$^2$/μg for the Fil-Tech electrodes, and $A$ is the surface area of the electrode which is 0.4987 cm$^2$.

*High Resolution Mass Spectrometry*

To accumulate mass on the electrode surface for further analysis, a Pt mesh electrode from BASi Research Products with an approximate surface area of 5 cm$^2$ was used as the working electrode with a Pt wire counter electrode and an Ag/AgCl (sat. KCl) reference electrode. The CV scan window was shortened to avoid the oxidation wave where most of the accumulated mass desorbed from the electrode surface. CV measurements were run in as-brewed coffee with about 1.5 %TDS for 50-100 cycles. Immediately upon CV cycling completion, the working electrode was removed from the sample, patted dry with a Kimwipe, and submerged into a solution of 80% 18 MΩ·cm water and 20% HPLC grade acetonitrile (Sigma-Aldrich) (v/v). The solution and working electrode were sonicated for 10 minutes and then the electrode was removed and rinsed. This process was repeated a total of four times to ensure that the solution contained enough redissolved residue for mass spectrometry analysis.

A Thermo Scientific Vanquish UPLC paired with a Q-Exactive tandem orbitrap mass spectrometer, operating in positive polarity mode with a nominal resolution of 70,000, was used to identify unknown compounds and to quantify caffeine in the residue sample. An ACE Excel 2 SuperC18 UPLC column (100mm x 2.10mm, 2um) was used to achieve separation of the compounds in the residue using a mobile phase consisting of: ultrapure water, methanol (Honeywell), acetonitrile (JT Baker), and 1M acetic acid. All mobile phase components also contained 0.1% formic acid. The LC flow rate was 350 μL/min, and the gradient used for the compound identification runs can be found in **Table S3**. The MS system was run in full MS mode (over a range of 150 – 1000 m/z) for compound identification. Identification of caffeine was confirmed by investigating the TIC peak located at the same retention time as that of a caffeine standard solution and matching the highest intensity peak in the mass spectrum at that retention time to that of protonated caffeine to less than 2ppm of mass error. Further, there was excellent agreement between the observed isotope pattern and a simulation of the expected pattern for caffeine generated using the Freestyle data analysis software package (Thermo Scientific), **Figure S8**.

Quantification of caffeine was performed using targeted selected ion monitoring (t-SIM) mode centered at 195.0876 m/z with an isolation width of 1.0 m/z. The unknown, and each standard, was measured in triplicate and the data processed using an automated peak-finding and peak-fitting algorithm available in the Freestyle data analysis application. Each sample was injected using the UPLC and a short mobile phase gradient was used with a total run time of 8 minutes per injection. The details of this gradient can be found in **Table S4**.

*Roasting*

A green coffee, El Tambo from Cauca, Colombia, was sourced from Tailored Coffee Roasters, Eugene, OR. This coffee is representative of an average specialty coffee, and costs approximately $10/kg. The coffee was roasted in 50 g sample batches using an IKAWA Pro50. Sample batches were roasted using a simple roast profile and the associated profiles are presented in **Figure 3a**. The initial roast profile (roast 1a) went for 6 minutes and 17 seconds and ended at 211.11 °C (412 °F). Additional roasts were each extended by an additional 30 seconds

and 1.1 °C (2 °F) for a total of 8 minutes and 17 seconds and an ending temperature of 215.56 °C (420 °F) for roast 1e. The roast extension was doubled to 60 seconds and 2.2 °C (4 °F) between roasts 1e and 1f to ensure a sufficiently dark sample was included. Roast colors are presented on the Agtron Gourmet scale, where Agtron values range from 0 (black) to 150 (green, unroasted). The values were measured using the Coffee Dipper spectrophotometer in whole bean mode. The data is collated in **Table S4**.

*Sourced coffee samples*

Coffee samples used in **Figure 4** were sourced from Colonna Coffee, Bath, UK. Those coffees were roasted to the same specifications with terminal whole bean colors of 98.9, 93.9, 92.8, and 93.6 mapping to samples 1, 2, 3, and 4, respectively. Sample 1 was rejected on their cupping table while the other three samples were accepted.

*Computational models*

All calculations were performed using the Vienna Ab initio Simulation Package (VASP 5.4.4) [67], [68] using a plane wave basis set and projector augmented-wave pseudopotentials.[69] Bare surfaces of Pt(100), Pt(110), and Pt(111) as well as gas-phase molecules of caffeine and 5-CQA were geometrically optimized with density functional theory using a cutoff energy of 400 eV, the PBEsol functional,[70] and a k-point spacing of 0.03 2π/Å with a 20 Å vacuum layer. Interlayer van der Waals dispersion interactions were modeled using the DFT+D3 formalism with Becke-Johnson damping.[71], [72] The unit cells of Pt(*hkl*) contained 4 discrete layers of Pt atoms, and the ionic coordinates of the bottom 2 layers were frozen as bulk Pt while the top 2 layers were allowed to relax. DFT optimizations of caffeine/Pt(*hkl*) and chlorogenic acid/Pt(*hkl*) were initialized from minimum energy configurations derived from molecular dynamics simulations and used a single k-point at Gamma. For caffeine/Pt(*hkl*), 4 layers of Pt were used with the bottom 2 layers frozen, and for chlorogenic acid/Pt(*hkl*), 2 layers of Pt were used with the bottom layer frozen. Molecular dynamics simulations were performed with a 300 eV cutoff energy using the Nosé-Hoover thermostat[73], [74] with 1 fs time steps beginning at 300 K and approaching 1 K, based on prior research on glycerol adsorption at Pt surfaces.[75] For both DFT and MD, the convergence criteria were 10 meV/Å for atomic forces and $1 \times 10^{-6}$ eV for total energy.

**Data availability**

All data generated in this work is contained within the article and Supplemental Information. All raw data is available from the corresponding author upon request.

**Acknowledgements**

This work was supported by the Coffee Science Foundation, underwritten by Nuova Simonelli. The authors are also grateful for the support from the National Science Foundation under grant no. 2237345 and the Camille and Henry Dreyfus Foundation. We are grateful to Dr. Jiawei Huang for his help performing QCM experiments. The authors also appreciate many helpful discussions with Dr. Andreas Stonas and Brian Sung. We are grateful to Maxwell Dashwood at Colonna

Coffee for providing the four samples of roasted coffee. Other coffee used in this study inevitably came from in-house roasting but the methods were developed using coffees donated to the laboratory by Counter Culture Coffee, Onyx Coffee Lab, Sweetbloom, Moongoat, Passenger, Black and White, Tim Wendelboe, Monogram, Archetype, Farmers Union, Tailored, Reverie, Blueprint, ONA, Phil and Sebastian, Proud Mary, Keurig Dr. Pepper, Mostra, Coffea Circulor, Mother Tongue, Creature, and Intelligentsia.

**Author contributions**

The project was conceived by C.H.H. R.E.B., D.L.P., and L.C.W. performed electrochemical measurements. E.J.R. collected the particle size distributions. Chromatography and mass spectrometry was performed by J.R.W. Statistical analysis was performed by B.J.A. All authors contributed to writing the manuscript.

**Competing interests**

C.H.H. and D.L.P. have a financial interest in The Overpotential Company, a start-up company that is seeking to commercialize electrochemically modified food. The remaining authors declare no competing interests.

# An Electrochemical Appraisal of Coffee Qualities

Robin E. Bumbaugh[a,†], Doran L. Pennington[a,†], Lena C. Wehn[a], Elias J. Rheingold[a], Joshua R. Williams[b], Benjamín J. Alemán[c], Christopher H. Hendon[a]*

[a] Department of Chemistry and Biochemistry, University of Oregon, Eugene, OR, 97403, USA *email: chendon@uoregon.edu

[b] Department of Chemistry, Drexel University, Philadelphia, PA, 19104, USA

[c] Department of Physics, University of Oregon, Eugene, OR, 97403, USA

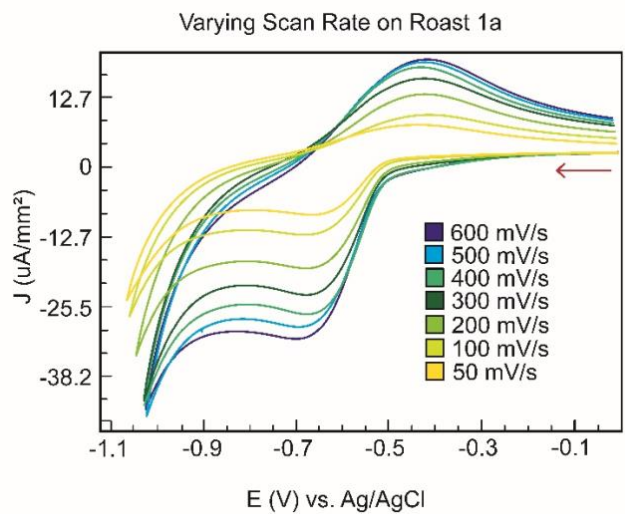

**Figure S1.** Scan rate dependence of HUPD peaks in undiluted brewed coffee. Reduction and oxidation peak centers do not correlate linearly with log base 10 of the scan rate, indicating non-Nernstian behavior.

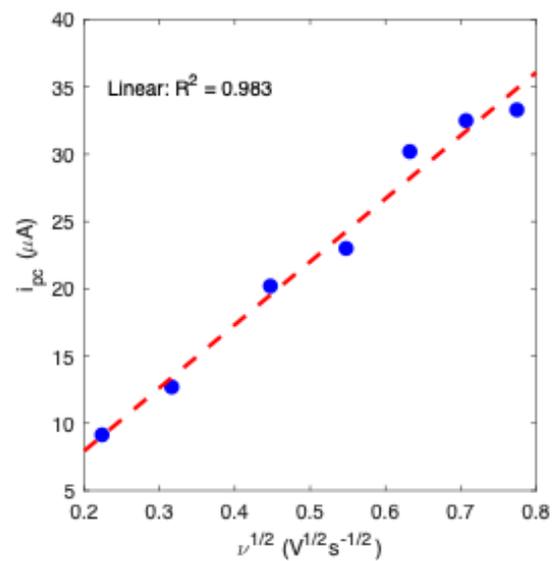

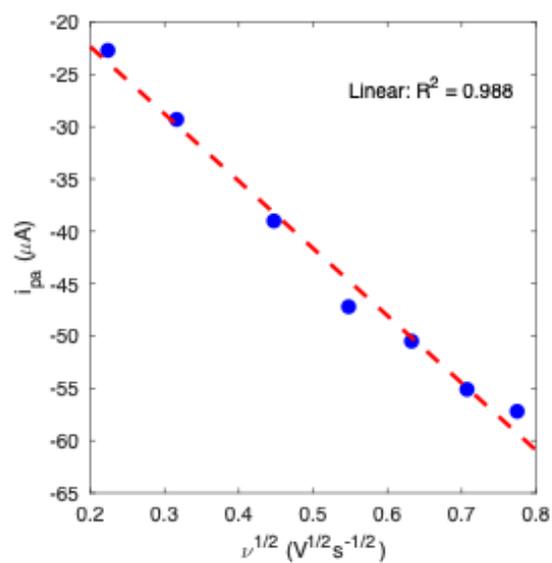

**Figure S2.** Linear dependence of peak current on the square root of the scan rate for cathodic ($i_{pc}$) and anodic ($i_{pa}$) features near -0.6 V corresponding to a diffusion-controlled process, from 50 to 600 mV/s.

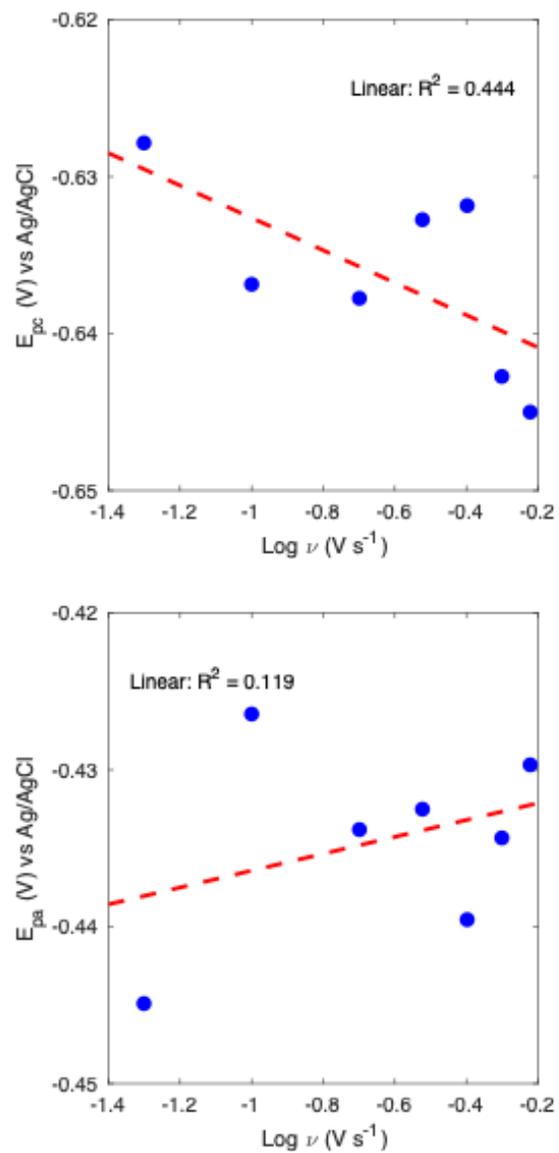

**Figure S3.** Lack of a linear increase in peak separation with increasing scan rate for cathodic ($E_{pc}$) and anodic ($E_{pa}$) features near -0.6 V indicative of a reversible nernstian process, from 50 to 600 mV/s.

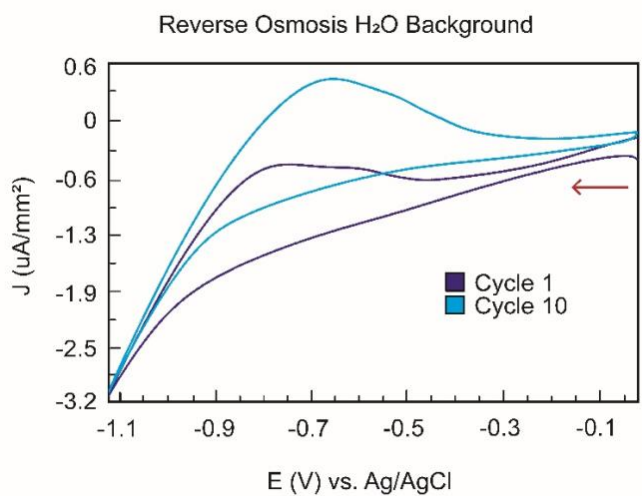

**Figure S4.** Background CV of water purified using the Pentair Everpure Conserv 75E Reverse Osmosis system. Scan rate of 200 mV/s.

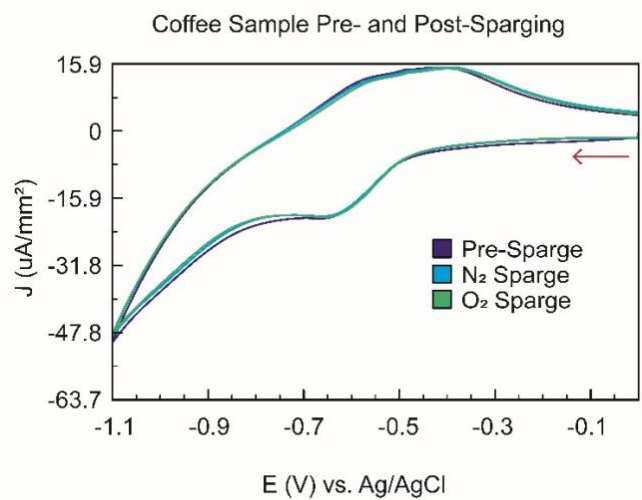

**Figure S5.** Overlay of the first CV cycle of brewed coffee with and without sparging. Dark blue trace is the brewed coffee prior to any sparging; light blue trace is following 20 minutes of sparging with $N_2$; green trace is following an additional 20 minutes of sparging with $O_2$. Scan rate 200 mV/s.

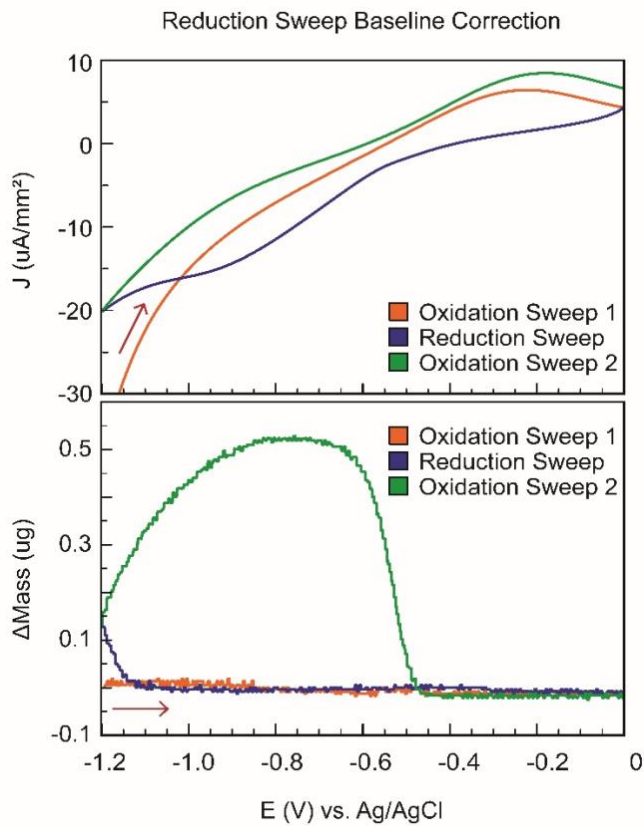

**Figure S6.** CV and associated QCM mass data in coffee at 1.36 %TDS. Beggining at -1.2 V vs. Ag/AgCl and scanned oxidatively, as opposed to starting at 0 V and scanning reductively to start. QCM shows no mass gain during the initial oxidative sweep (orange trace). Mass is only accumulated following at potentials negative of $H_{UPD}$, during the first reductive sweep (blue trace).

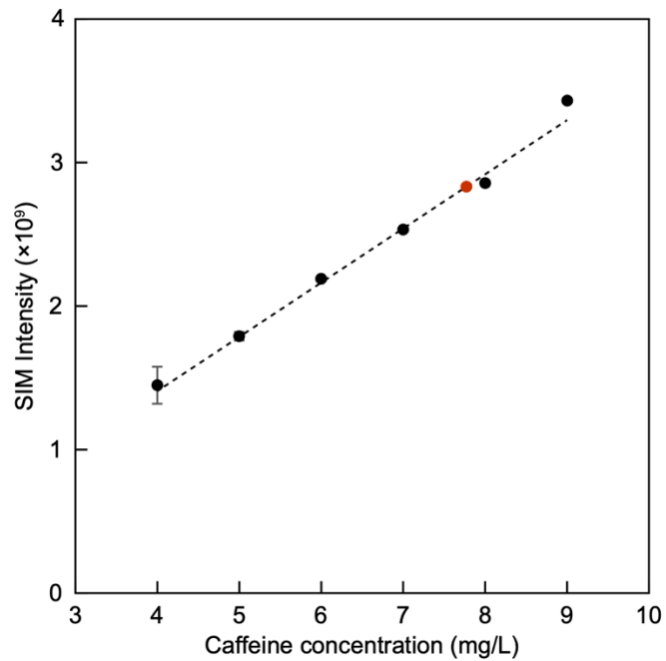

**Figure S7**. Calibration curve for quantification of caffeine concentration. Error bars are included from pentaplicate runs. The unknown sample obtained from electrochemical cycling is shown in red.

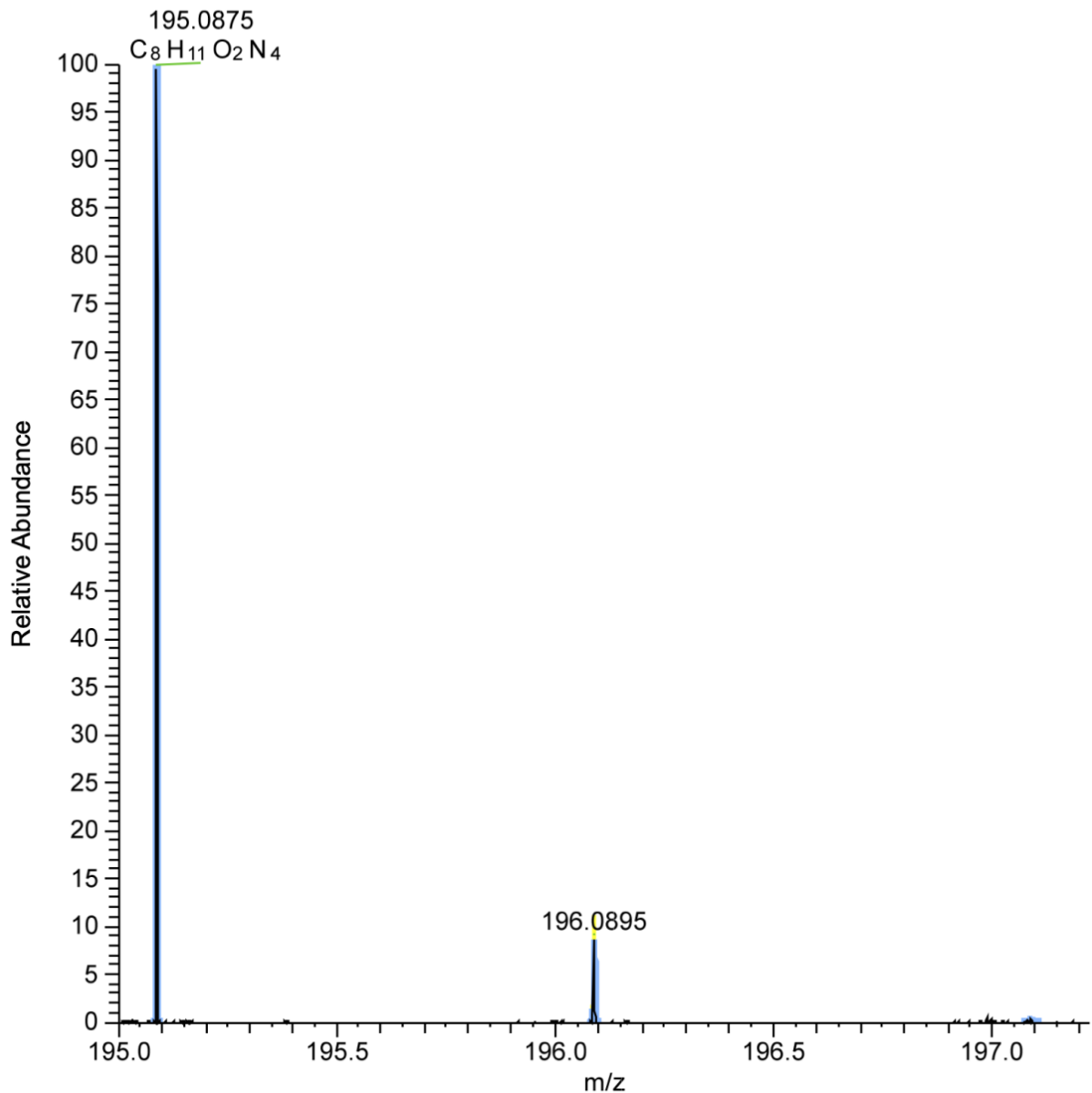

**Figure S8**. High resolution mass spectrum of the chromatographic peak at a retention time of 4.70min, consistent with that of caffeine under the chromatographic conditions used. The monoisotopic peak (195.0875 m/z) matches that expected from protonated caffeine (195.08765 m/z) to a mass error of less than 1ppm. The blue overlay is a simulation of the isotope pattern expected for a compound with the formula of protonated caffeine.

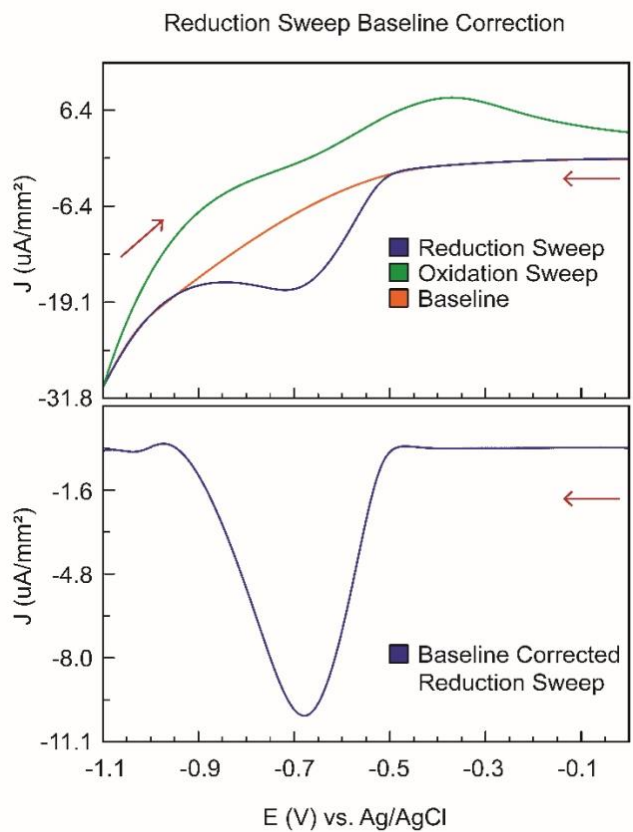

**Figure S9.** An example of background subtraction performed in OriginPro9 to integrate reductive feature. The top panel shows the first CV curve of roast 1a at 1.56 %TDS. The bottom panel shows the reduction sweep of the CV after the baseline correction was subtracted.

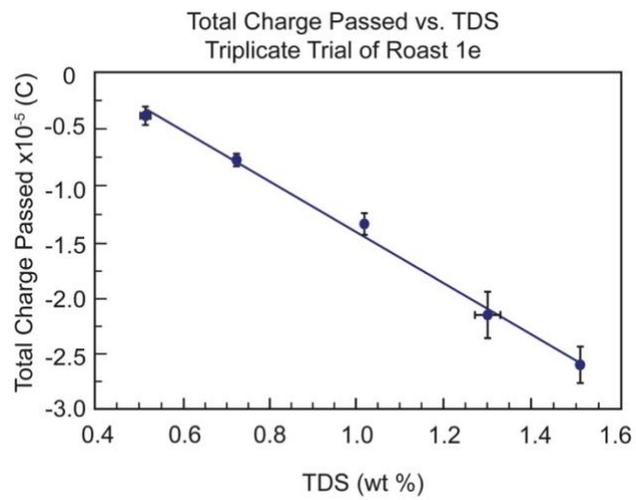

**Figure S10.** Total charge passed versus %TDS performed in triplicate using Roast 1e with associated error bars.

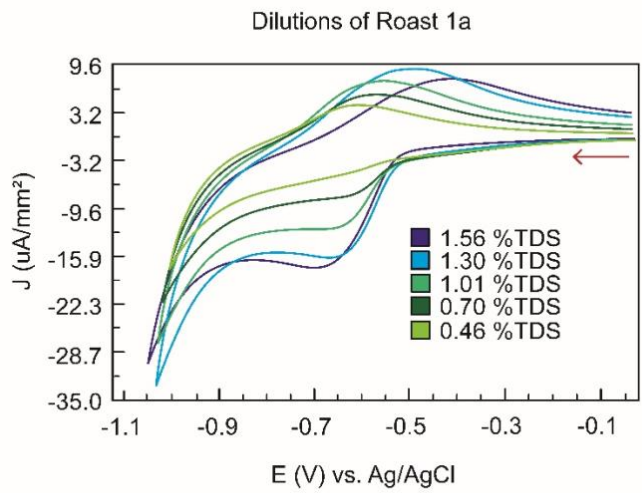

**Figure S11.** Overlay of the first CV cycle of Roast 1a via cupping brew method at multiple %TDS values. Scan rate 200 mV/s.

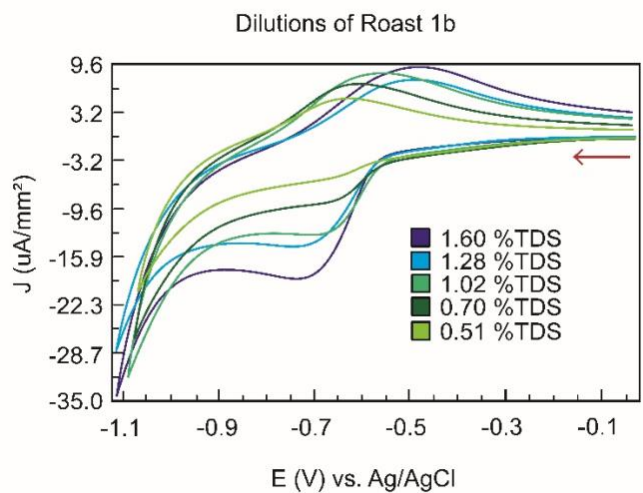

**Figure S12.** Overlay of the first CV cycle of Roast 1b via cupping brew method at multiple %TDS values. Scan rate 200 mV/s.

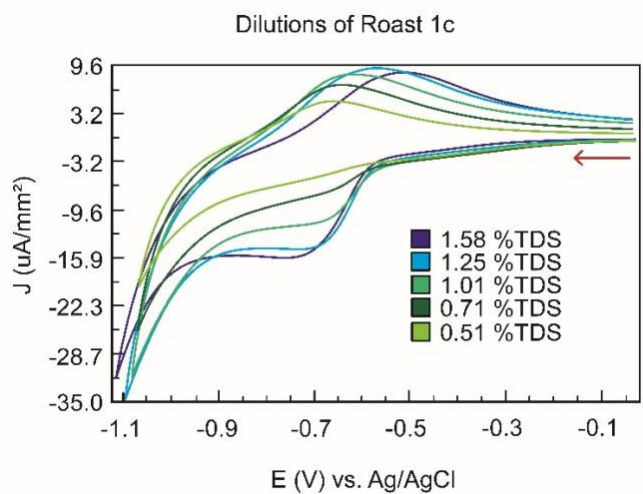

**Figure S13.** Overlay of the first CV cycle of Roast 1c via cupping brew method at multiple %TDS values. Scan rate 200 mV/s.

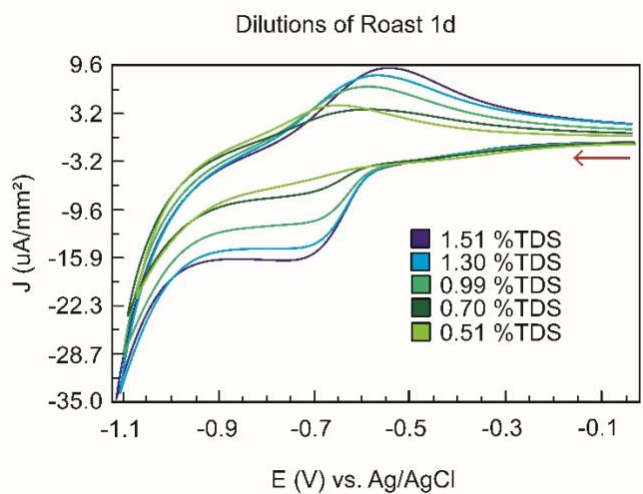

**Figure S14.** Overlay of the first CV cycle of Roast 1d via cupping brew method at multiple %TDS values. Scan rate 200 mV/s.

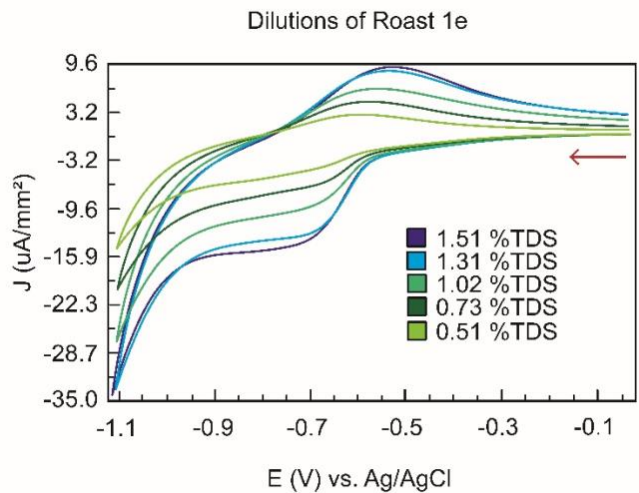

**Figure S15.** Overlay of the first CV cycle of Roast 1e via cupping brew method at multiple %TDS values. Scan rate 200 mV/s.

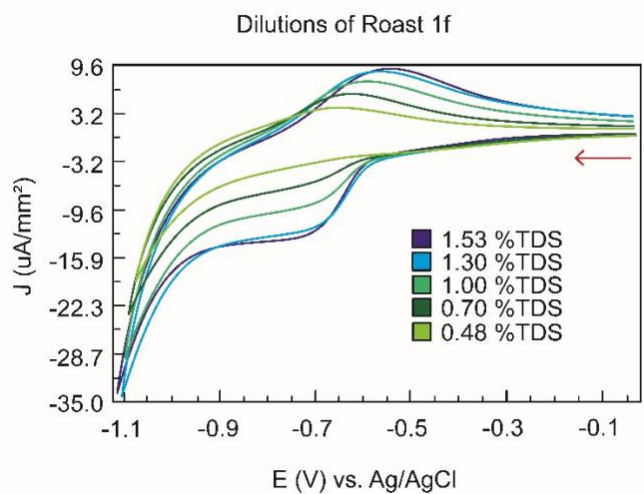

**Figure S16.** Overlay of the first CV cycle of Roast 1f via cupping brew method at multiple %TDS values. Scan rate 200 mV/s.

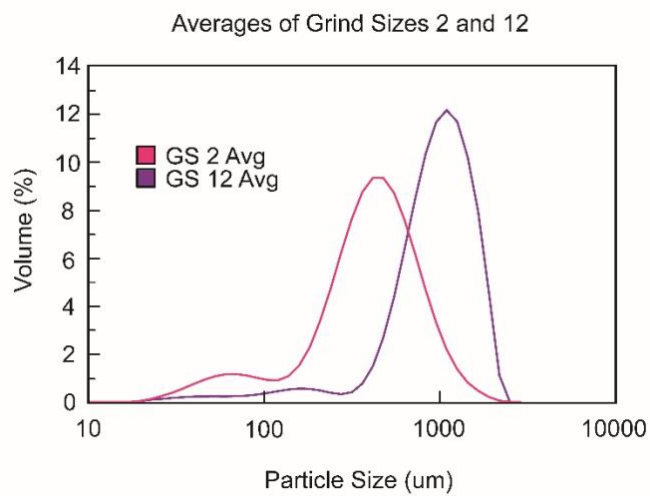

**Figure S17**. Particle size distributions at grinder setting 2 and 12. Setting 2 was used for ground bean color assessment, while setting 12 was used for coffee cupping brews.

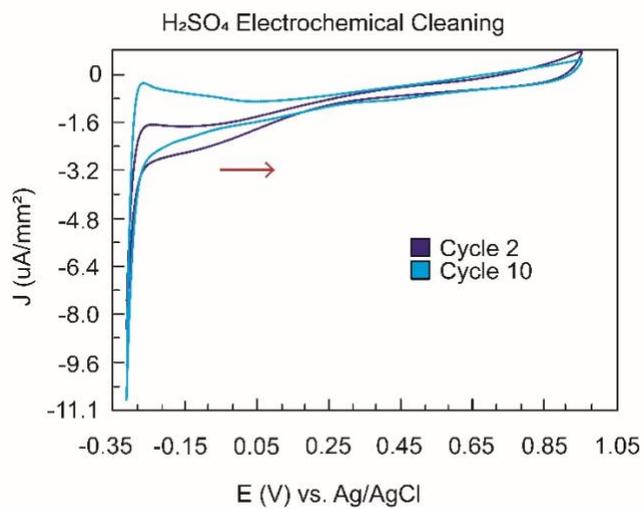

**Figure S18.** Example of electrochemical cleaning of platinum working electrode in 50 mM H₂SO₄. Cycles were repeated until the features visibly overlaid with one another. Scan rate 300 mV/s.

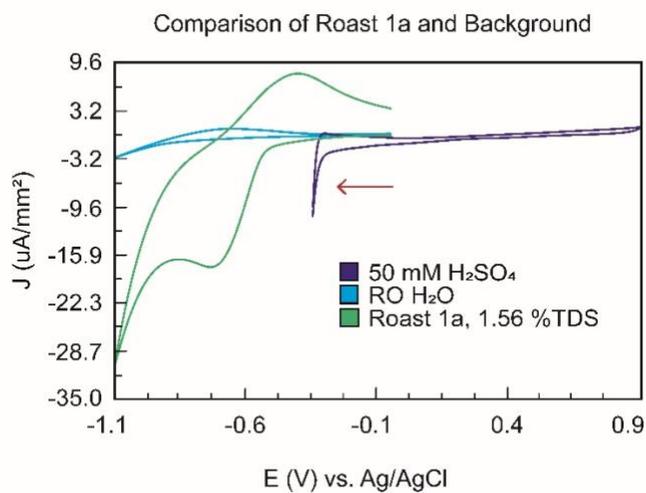

**Figure S19.** Overlay of Roast 1a at 1.56 %TDS with a background scan of the reverse osmosis $H_2O$ used for coffee brewing and 50 mM solution of $H_2SO_4$ used for electrochemical polishing of electrodes. Scan rates for $H_2O$ and coffee were 200 mV/s and 300 mV/s for $H_2SO_4$.

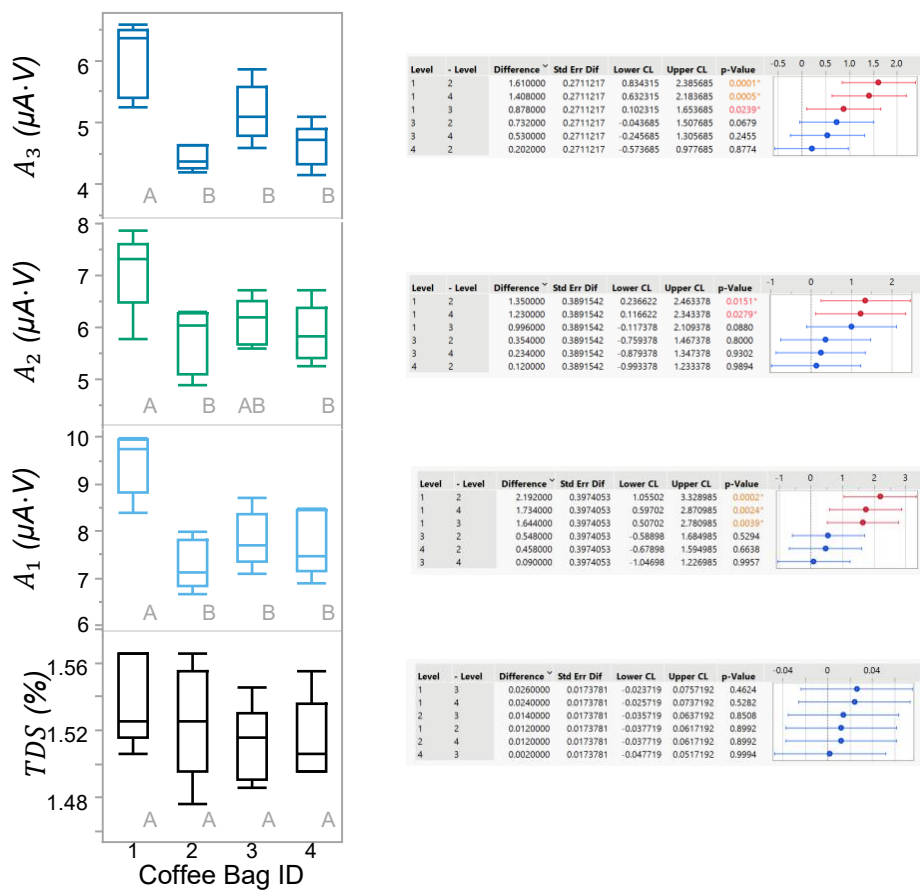

**Figure S20.** $TDS$ and peak integrated area distributions and Tukey HSD ANOVA analysis for coffee roast samples. (Lower panel) The measured $TDS$ was statistically identical across all four samples (connecting letters A class); TDS differences (insignificant) ranged from $0.002 \pm 0.02$ to $0.03 \pm 0.02$, with $P = 0.999$ to $P = 0.462$, respectively. (Second panel from bottom) The measured peak integrated area for the first cycle, $A_1$, was split into two distinct classes—A and B—with coffee bag ID 1 in class A and ID's 2-4 in class B. The differences between classes A and B were $2.19 \pm 0.40$ µA·V (1-2; $P = 0.0002$), $1.73 \pm 0.40$ µA·V (1-4; $P = 0.0024$), and $1.64 \pm 0.40$ µA·V (1-3; $P = 0.0039$). Differences between treatments in the B class (insignificant), ranged from 0.09 to 0.55 µA·V ($P = 0.996$ to $P = 0.529$, respectively). (Second panel from top) The measured peak integrated area for the second cycle, $A_2$, was split into two distinct classes—A and B—with coffee bag ID 1 in class A, ID's 2,4 in class B, and ID 3 in both classes A and B. The differences between classes ID's 1 and 2,4 were 1.35 µA·V (1-2; $P = 0.015$) and 1.23 µA·V (1-4; $P = 0.0039$); the difference between ID 1 and 3 was 0.996 µA·V but only at $P = 0.088$ (i.e. close the $\alpha = 0.05$ threshold). Differences between treatments in the B class (insignificant), ranged from $0.12 \pm 0.39$ to $0.35 \pm 0.39$ µA·V ($P = 0.989$ to $P = 0.800$, respectively). (Top panel) The measured peak integrated area for the third cycle, $A_3$, was split into two distinct classes—A and B—with coffee bag ID 1 in class A, ID's 2-4 in class B. The differences between classes ID's 1 and 2-4 were $1.61 \pm 0.27$ µA·V (1-2; $P = 0.0001$), $1.41 \pm 0.27$ µA·V (1-4; $P = 0.0005$), and $0.88 \pm 0.27$ µA·V (1-3; $P = 0.0239$). Differences between treatments in the B class (insignificant), ranged from 0.20 to 0.73 µA·V ($P = 0.877$ to $P = 0.07$, respectively); the difference between ID's 2 and 3 (0.73 µA·V, $P = 0.07$) was close to the $\alpha = 0.05$ threshold.

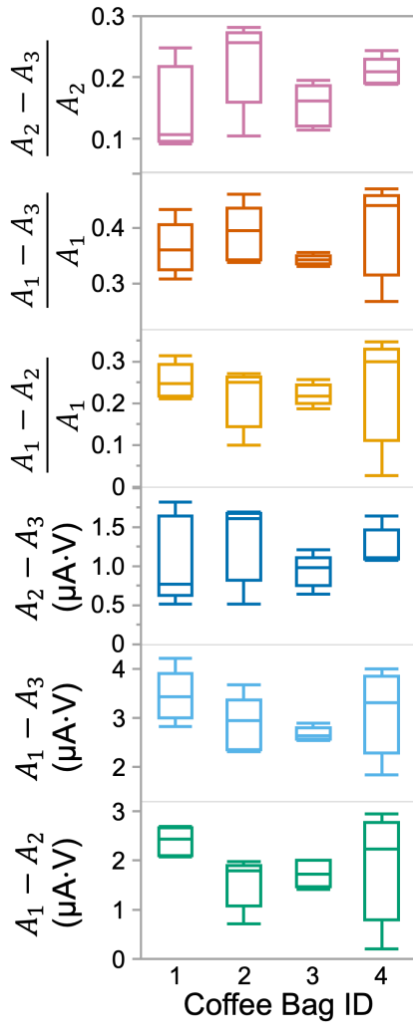

| Y | Level | - Level | Diff. | Std Err Dif | Lower CL | Upper CL | p-value |
|---|---|---|---|---|---|---|---|
| A1-A2 | 1 | 2 | 0.842 | 0.406 | -0.319 | 2.003 | 0.203 |
| A1-A2 | 1 | 3 | 0.648 | 0.406 | -0.513 | 1.809 | 0.408 |
| A1-A2 | 1 | 4 | 0.504 | 0.406 | -0.657 | 1.665 | 0.61 |
| A1-A2 | 4 | 2 | 0.338 | 0.406 | -0.823 | 1.499 | 0.838 |
| A1-A2 | 3 | 2 | 0.194 | 0.406 | -0.967 | 1.355 | 0.963 |
| A1-A2 | 4 | 3 | 0.144 | 0.406 | -1.017 | 1.305 | 0.984 |
| A1-A3 | 1 | 3 | 0.766 | 0.372 | -0.298 | 1.83 | 0.209 |
| A1-A3 | 1 | 2 | 0.582 | 0.372 | -0.482 | 1.646 | 0.425 |
| A1-A3 | 4 | 3 | 0.44 | 0.372 | -0.624 | 1.504 | 0.646 |
| A1-A3 | 1 | 4 | 0.326 | 0.372 | -0.738 | 1.39 | 0.817 |
| A1-A3 | 4 | 2 | 0.256 | 0.372 | -0.808 | 1.32 | 0.9 |
| A1-A3 | 2 | 3 | 0.184 | 0.372 | -0.88 | 1.248 | 0.959 |
| A2-A3 | 2 | 3 | 0.378 | 0.259 | -0.362 | 1.118 | 0.482 |
| A2-A3 | 4 | 3 | 0.296 | 0.259 | -0.444 | 1.036 | 0.668 |
| A2-A3 | 2 | 1 | 0.26 | 0.259 | -0.48 | 1 | 0.748 |
| A2-A3 | 4 | 1 | 0.178 | 0.259 | -0.562 | 0.918 | 0.9 |
| A2-A3 | 1 | 3 | 0.118 | 0.259 | -0.622 | 0.858 | 0.967 |
| A2-A3 | 2 | 4 | 0.082 | 0.259 | -0.658 | 0.822 | 0.989 |
| (A1-A2)/A1 | 1 | 2 | 0.042 | 0.049 | -0.099 | 0.182 | 0.832 |
| (A1-A2)/A1 | 1 | 3 | 0.032 | 0.049 | -0.109 | 0.172 | 0.915 |
| (A1-A2)/A1 | 4 | 2 | 0.023 | 0.049 | -0.117 | 0.164 | 0.963 |
| (A1-A2)/A1 | 1 | 4 | 0.018 | 0.049 | -0.122 | 0.158 | 0.982 |
| (A1-A2)/A1 | 4 | 3 | 0.014 | 0.049 | -0.127 | 0.154 | 0.992 |
| (A1-A2)/A1 | 3 | 2 | 0.01 | 0.049 | -0.13 | 0.15 | 0.997 |
| (A1-A3)/A1 | 4 | 3 | 0.055 | 0.035 | -0.044 | 0.154 | 0.412 |
| (A1-A3)/A1 | 2 | 3 | 0.048 | 0.035 | -0.051 | 0.147 | 0.529 |
| (A1-A3)/A1 | 4 | 1 | 0.034 | 0.035 | -0.065 | 0.134 | 0.756 |
| (A1-A3)/A1 | 2 | 1 | 0.027 | 0.035 | -0.072 | 0.126 | 0.861 |
| (A1-A3)/A1 | 1 | 3 | 0.021 | 0.035 | -0.079 | 0.12 | 0.932 |
| (A1-A3)/A1 | 4 | 2 | 0.007 | 0.035 | -0.092 | 0.107 | 0.997 |
| (A2-A3)/A2 | 2 | 1 | 0.078 | 0.034 | -0.019 | 0.175 | 0.139 |
| (A2-A3)/A2 | 2 | 3 | 0.069 | 0.034 | -0.028 | 0.166 | 0.212 |
| (A2-A3)/A2 | 4 | 1 | 0.064 | 0.034 | -0.033 | 0.161 | 0.274 |
| (A2-A3)/A2 | 4 | 3 | 0.055 | 0.034 | -0.042 | 0.152 | 0.393 |
| (A2-A3)/A2 | 2 | 4 | 0.014 | 0.034 | -0.083 | 0.111 | 0.974 |
| (A2-A3)/A2 | 3 | 1 | 0.009 | 0.034 | -0.088 | 0.106 | 0.994 |

**Figure S21**. Tukey HSD ANOVA analysis of differences between scans 1, 2, and 3, both normalized (divided by the scan area) and native differences between the four Colonna coffees provided for Figure 4. Large *p*-values for all deltas/normalized deltas indicates that we retain the null hypothesis that the deltas/normalized deltas indistinguishable.

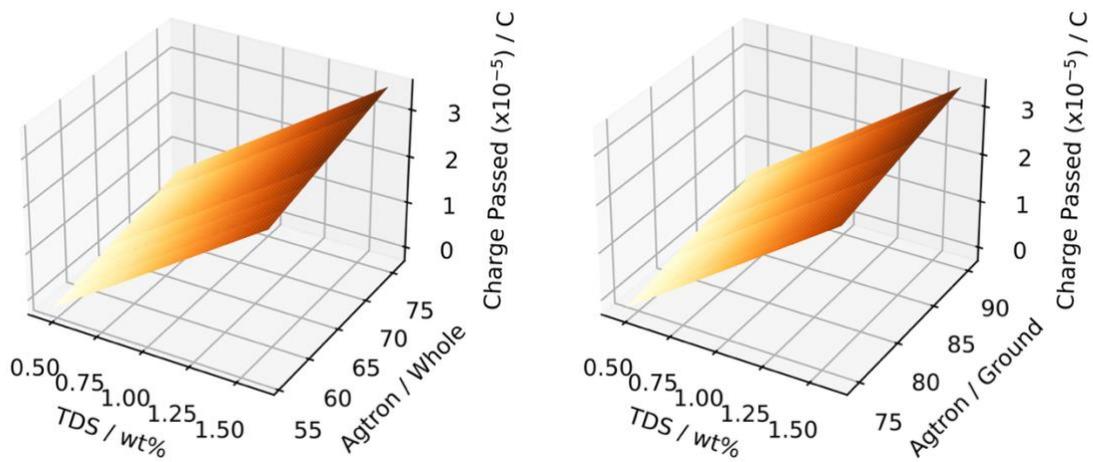

**Figure S22.** The electrochemical brew plane expressed as whole bean and ground Agrton colors.

**Table S1.** %TDS, resistance, and pH values for coffee samples used for cyclic voltammetry.

| Roast | Sample | TDS | Resistance (Ω) | pH |
|---|---|---|---|---|
| 1a | Pure | 1.56 | 740.94 | 4.96 |
| 1a | Dil1 | 1.30 | 837.57 | 4.99 |
| 1a | Dil2 | 1.01 | 1052.93 | 5.04 |
| 1a | Dil3 | 0.70 | 1460.61 | 5.12 |
| 1a | Dil4 | 0.46 | 2091.03 | 5.24 |
| 1b | Pure | 1.60 | 733.99 | 5.06 |
| 1b | Dil1 | 1.28 | 875.55 | 5.08 |
| 1b | Dil2 | 1.02 | 1048.32 | 5.12 |
| 1b | Dil3 | 0.70 | 1425.42 | 5.19 |
| 1b | Dil4 | 0.51 | 1977.29 | 5.25 |
| 1c | Pure | 1.58 | 719.85 | 5.12 |
| 1c | Dil1 | 1.25 | 835.85 | 5.13 |
| 1c | Dil2 | 1.01 | 1048.65 | 5.17 |
| 1c | Dil3 | 0.71 | 1445.79 | 5.23 |
| 1c | Dil4 | 0.51 | 1922.64 | 5.31 |
| 1d | Pure | 1.51 | 768.58 | 5.17 |
| 1d | Dil1 | 1.30 | 876.40 | 5.19 |
| 1d | Dil2 | 0.99 | 1103.23 | 5.23 |
| 1d | Dil3 | 0.70 | 1507.73 | 5.30 |
| 1d | Dil4 | 0.51 | 1994.35 | 5.38 |
| 1e Run 1 | Pure | 1.50 | 789.60 | 5.30 |
| 1e Run 1 | Dil1 | 1.26 | 861.16 | 5.29 |
| 1e Run 1 | Dil2 | 1.01 | 1091.62 | 5.35 |
| 1e Run 1 | Dil3 | 0.72 | 1491.01 | 5.42 |
| 1e Run 1 | Dil4 | 0.51 | 1992.82 | 5.50 |
| 1e Run 2 | Pure | 1.50 | 786.07 | 5.30 |
| 1e Run 2 | Dil1 | 1.31 | 865.86 | 5.29 |
| 1e Run 2 | Dil2 | 1.01 | 1090.67 | 5.36 |
| 1e Run 2 | Dil3 | 0.72 | 1495.26 | 5.42 |
| 1e Run 2 | Dil4 | 0.53 | 1980.11 | 5.50 |
| 1e Run 3 | Pure | 1.51 | 771.35 | 5.30 |
| 1e Run 3 | Dil1 | 1.31 | 872.56 | 5.29 |
| 1e Run 3 | Dil2 | 1.02 | 1091.54 | 5.36 |
| 1e Run 3 | Dil3 | 0.73 | 1489.70 | 5.42 |
| 1e Run 3 | Dil4 | 0.51 | 1990.33 | 5.50 |
| 1f | Pure | 1.53 | 783.20 | 5.38 |
| 1f | Dil1 | 1.30 | 860.28 | 5.39 |
| 1f | Dil2 | 1.00 | 1077.01 | 5.44 |
| 1f | Dil3 | 0.70 | 1465.19 | 5.52 |

**Table S2.** Roasting conditions and their associated Agtron numbers. Grind setting 2 was used on an EK-43 to obtain the ground Agtron values.

| Roast | Total Time (min) | Final Temperature (°F) | Whole bean Agtron | Ground coffee Agtron |
|---|---|---|---|---|
| **1a** | 6:17 | 412 | 75.8 | 90.3 |
| **1b** | 6:47 | 414 | 72.2 | 88.8 |
| **1c** | 7:17 | 416 | 64.3 | 83.6 |
| **1d** | 7:47 | 418 | 61.2 | 80.3 |
| **1e** | 8:17 | 420 | 58.2 | 74.6 |
| **1f** | 9:17 | 424 | 55.7 | 76.1 |

**Table S3.** Liquid chromatography mobile phase gradient for compound identification analysis. All transitions are linear.

| Time (min) | Water (%) | Methanol (%) | Acetonitrile (%) | 1M Acetic Acid (%) |
|---|---|---|---|---|
| 0 | 67 | 8 | 5 | 20 |
| 4.5 | 63 | 12 | 5 | 20 |
| 28 | 0 | 55 | 40 | 5 |
| 35 | 0 | 55 | 40 | 5 |
| 40 (end) | 67 | 8 | 5 | 20 |

**Table S4.** Liquid chromatography mobile phase linear gradient for caffeine quantification. All transitions are linear.

| Time (min) | Water (%) | Methanol (%) | Acetonitrile (%) | 1M Acetic Acid (%) |
|---|---|---|---|---|
| 0 | 67 | 8 | 5 | 20 |
| 8 | 63 | 12 | 5 | 20 |